\documentclass[twocolumn]{webofc}
\usepackage[varg]{txfonts}   
\usepackage{booktabs}
\usepackage[utf8]{inputenc}
\usepackage{textcomp}
\usepackage{amsmath}
\usepackage[detect-all]{siunitx}
\usepackage{array} 
\newcolumntype{L}[1]{>{\raggedright\let\newline\\\arraybackslash\hspace{0pt}}m{#1}}
\newcolumntype{C}[1]{>{\centering\let\newline\\\arraybackslash\hspace{0pt}}m{#1}}
\newcolumntype{R}[1]{>{\raggedleft\let\newline\\\arraybackslash\hspace{0pt}}m{#1}}

\newcommand{\smax}{$S_{\mathrm{max}}$}

\DeclareSIUnit\mSv{\milli\sievert}
\DeclareSIUnit\mGy{\milli\gray}
\DeclareSIUnit\mrad{\milli\radian}
\DeclareSIUnit[number-unit-product = {}]\WET{WET}

\DeclareSIUnit[number-unit-product = ]\percent{\char`\%}

\graphicspath{{graphics/}{graphics/arch/}{Graphics/}{./}} 

\begin{document}
\title{Proton Tracking Algorithm in a Pixel-Based Range Telescope for Proton Computed Tomography}

\author{\firstname{Helge E. S.} \lastname{Pettersen}\inst{1,2,3}\fnsep\thanks{\email{helge.pettersen@helse-bergen.no}} \and
        \firstname{Ilker} \lastname{Meric}\inst{2}\and
        \firstname{Odd Harald} \lastname{Odland}\inst{1} \and 
        \firstname{Hesam} \lastname{Shafiee}\inst{2} \and 
        \firstname{Jarle R.} \lastname{Sølie}\inst{2} \and 
        \firstname{Dieter} \lastname{Röhrich}\inst{3}
      }

\institute{Department of Oncology and Medical Physics, Haukeland University Hospital, Norway
\and
           Department of Electrical Engineering, Western Norway University of Applied Sciences, Bergen, Norway
\and
           Department of Physics and Technology, University of Bergen, Norway
          }

\abstract{%
Proton Computed Tomography (CT) is a prototype imaging modality for the reconstruction of the Relative Stopping Power of a patient, for more accurate calculations of the dose distributions in proton therapy dose planning. The measurements needed for the reconstruction of a proton CT image are: \textit{i}) each initial proton vector incident on the imaged object, \textit{ii}) each proton vector incident on the front face of the detector and \textit{iii}) the stopping depth of each proton in the detector. In this study, a track reconstruction algorithm is adapted for a planned pixel-based particle-tracking range telescope for proton CT, called the Digital Tracking Calorimeter (DTC). The algorithm is based on the track-following scheme, in which a growing track searches for deeper-laying activated pixels, while minimizing the accumulated angular change. The algorithm is applied to Monte Carlo-simulated output of the DTC, showing that the DTC is able to reconstruct the tracks and find the depths of up to several hundred simultaneous proton tracks.
}
\maketitle
\section{Introduction}\label{sec:introduction}
In the recent decades, radiation therapy using charged particles (such as protons) has increased in usage as a treatment against cancer. This is partly due to the proliferation of such treatment modalities \cite{jermann-particle-2017}, and also due to the growing evidence that proton therapy enables superior dose distributions to and around the tumor areas, with the consequent possibility of reducing radiation-induced damages to healthy tissue in the proximity of the tumor \cite{dionisi-use-2014,leeman-proton-2017}.

An accurate dose calculation for proton therapy requires a precise knowledge of the protons' range in the patient, calculated from the tissue-specific Relative Stopping Power (RSP). The stoichiometric procedure \cite{schneider-calibration-1996}, in which the RSP is calculated from the attenuation of photons in the tissue, acquired using X-ray Computed Tomography (CT), has been shown to introduce proton range uncertainties in the order of 2\%--3\% \cite{paganetti-range-2012}. Dual Energy CT can further reduce these uncertainties by at least 0.4\% \cite{baer-potential-2017}.

A proposed imaging modality that may prove helpful in the reduction of the above-mentioned uncertainties is proton CT \cite{johnson-review-2018}. With a proton CT system, a high-energy proton beam\footnote{A high proton energy in this context means 230--330 MeV, depending on beam availability.} is directed at and traverses the patient. Each individual proton's residual energy (corresponding to that proton's Water Equivalent Path Length, or WEPL, through the patient) is measured after traversing the patient: either with a range telescope or with a scintillator-based calorimeter. Due to the Multiple Coulomb Scattering (MCS) of the protons inside the patient, a separate set of positional detectors are usually placed upstream and downstream relative to the patient, so that each proton's curved path can be estimated using Bayesian methods \cite{krah-2018-comprehensive}.

\begin{figure}
\includegraphics[width=\columnwidth]{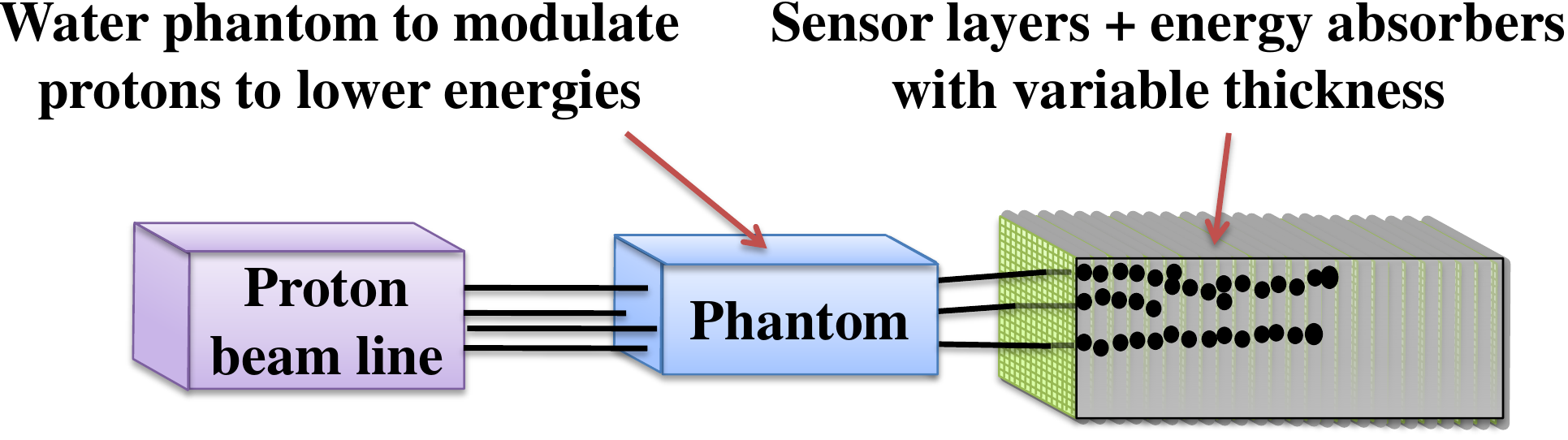}
\caption{The simulated setup of the DTC prototype, using a pencil beam and an energy-degrading water phantom \cite{pettersen-thesis-2018}.}\label{fig:dtc-setup}
\end{figure}

A three-dimensional RSP volume can then be reconstructed from the WEPL measurements using either iterative methods \cite{karbasi-incorporating-2015} or by modifications of Filtered Backprojection algorithms that allows for curved paths \cite{rit-filtered-2013}. The RSP accuracy of this procedure has been shown to be in the order of 1\%, significantly improving the range accuracy compared to existing methods \cite{r.-p.-johnson-fast-2016}.

Proton CT has not yet been clinically realized as an imaging modality. Several experimental setups have been developed \cite{johnson-review-2018}. Some of the challenges connected to proton CT includes the handling of the high proton beam intensities necessary to reduce image acquisition time, \textit{i.e.}\ requiring fast readout electronics; mitigating the effects of MCS through precise proton path estimations; and the realization of a compact detector system that can be introduced into the clinic.

\begin{figure}
\includegraphics[width=\columnwidth]{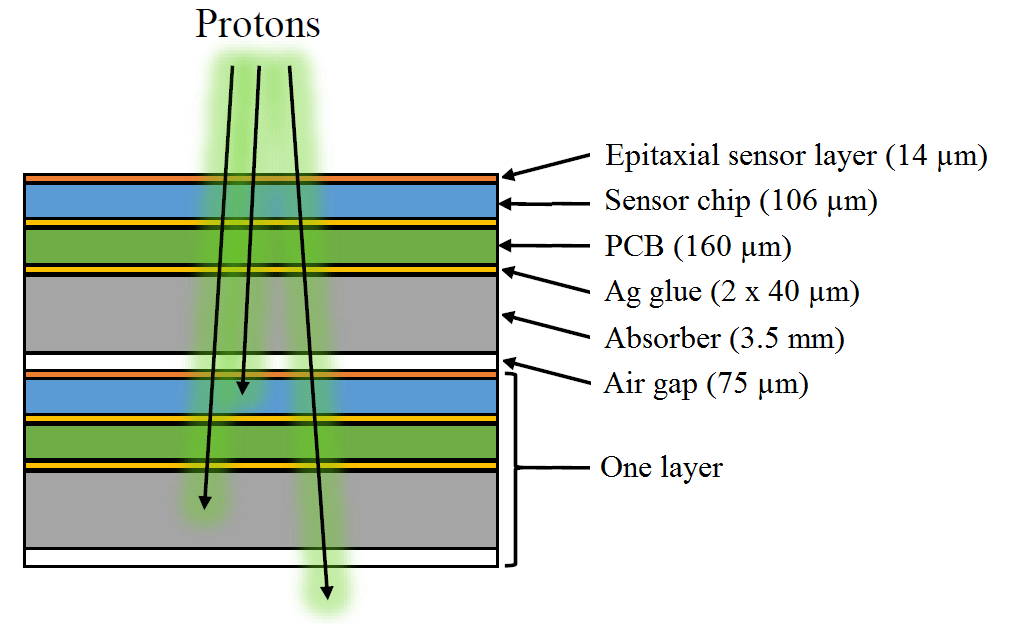}
\caption{A close-up of two layers, consisting of components that approximate the current planned layer design \cite{pettersen-thesis-2018}.}\label{fig:layer-setup}
\end{figure}

\subsection{The Digital Tracking Calorimeter}\label{sec:dtc}
Based on previous experience with a proof-of-concept pixel-based range telescope for proton CT \cite{pettersen-proton-2017}, hereafter a Digital Tracking Calorimeter (DTC), a second generation DTC is currently under development and construction. The schematics of the setup is shown in Fig.\ \ref{fig:dtc-setup}. The detector consists of approximately 40 layers, each layer a \SI{\sim 27 x 15}{\cm\squared} high-granularity 1-bit pixel sensor array using the ALPIDE chip \cite{mager-alpide-2016}, followed by an energy absorber of \SI{3.5}{\mm} aluminum.\footnote{The choice of absorber material and thickness has been made based on MC simulations of several possible geometries, followed by an evaluation of the range resolution and tracking quality in the respective geometries: see \cite{pettersen-thesis-2018} for more details.} The integration time of the chips can be customized, and the data acquisition hardware is expected to handle values of 5--10 µs. 

With the DTC, multiple proton tracks can be reconstructed from a single readout frame, \textit{i.e.}\ from a snapshot of the pixel values. The idea of applying in-detector tracking to increase the proton intensity capacity is not a new one \cite{esposito-cmos-2015}, but a complete setup has not yet been reported on. The DTC is planned to be used in conjunction with a Pencil Beam Scanning system, where a thin proton beam is scanned across the object to be imaged: the upstream proton vector can be estimated from the beam position (simplifying the proton CT setup), while the downstream proton vector is measured in the first pixel layers \cite{krah-2018-comprehensive}.

A high quality tracking algorithm maximizes the proton CT system capabilities in terms of contributing to increased particle rates, \textit{i.e.}\ a higher incident beam intensity. In this study, such a tracking algorithm is proposed, based on similar experiments (mainly in High Energy Physics \cite{amrouche-track-2017}) and prior experience with the DTC \cite{pettersen-proton-2017,pettersen-thesis-2018}.

\section{Methods}\label{sec:methods}

\begin{figure}
\includegraphics[width=\columnwidth]{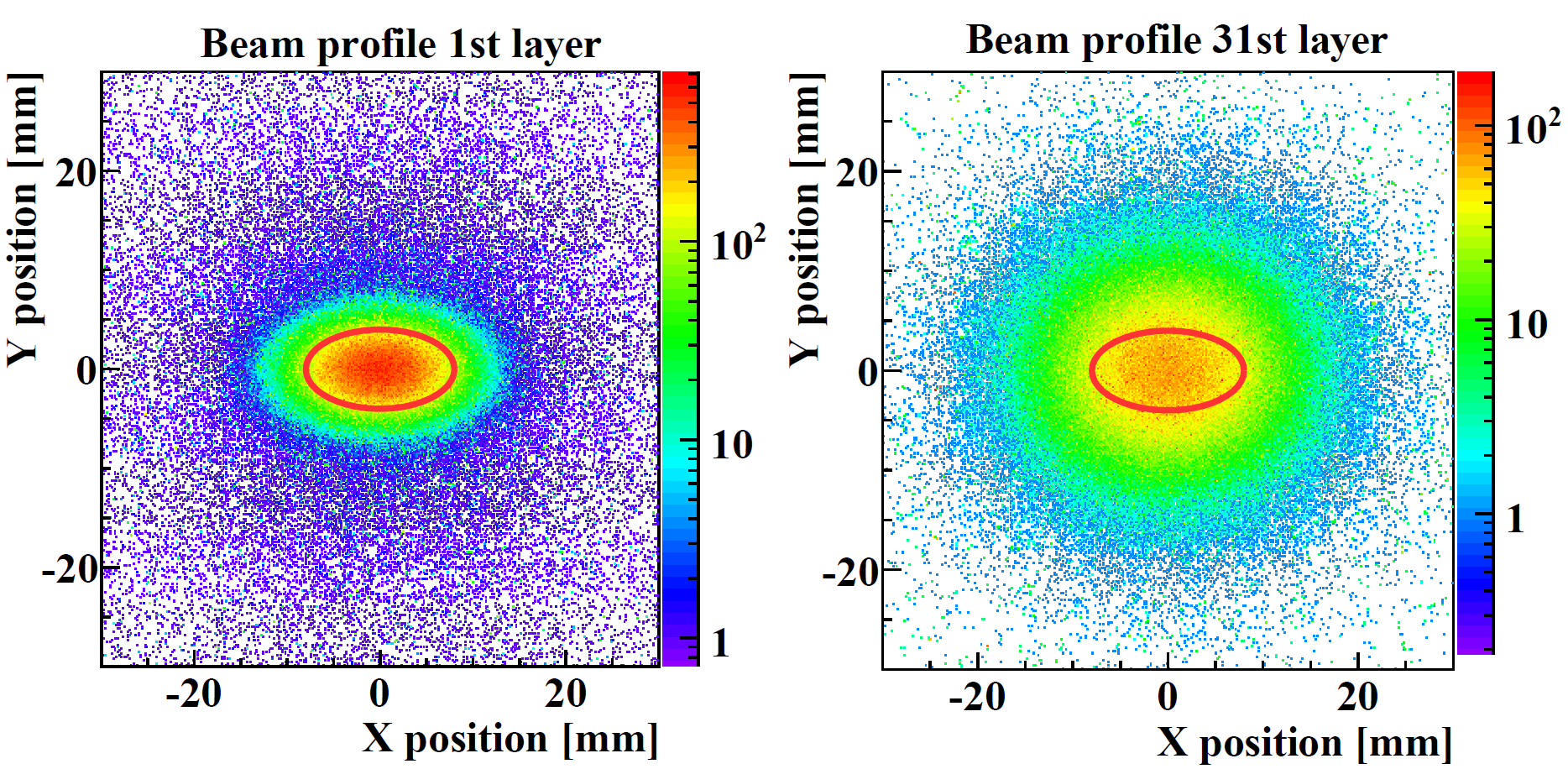}
\caption{Lateral profile of a pencil beam with spot size $\sigma_{xy} = (\SI{4}{\mm}, \SI{2}{\mm})$. The initial $2\sigma_{xy}$ ellipse is shown in red \cite{pettersen-thesis-2018}.}\label{fig:protonbeam}
\end{figure}

\subsection{The Monte Carlo simulations}\label{sec:montecarlo}
As the DTC is currently under construction, the development of a suitable tracking algorithm is performed on results from Monte Carlo (MC) simulations.

The Geant4-based MC program GATE version 8.1 \cite{jan-gate:-2004} has been applied, using with the physics builder list \texttt{QGSP\_BIC\_EMZ} with the corresponding default values for particle production threshold and variable transport step sizes: the maximum step sizes decrease towards the \textit{Bragg Peak} area of high energy deposition close to the protons' range. the above settings have previously been suggested for simulations in particle therapy \cite{grevillot-optimization-2010}.

The sensor chip is modeled as in Fig.\ \ref{fig:layer-setup}, with consecutive slabs of the following: a 14 µm active (epitaxial) Si layer, a 106 µm passive Si layer, glued (with Ag glue) to a 160 µm thick PCB layer, glued to the aluminum absorber, followed by a 75 µm air gap. While this is a simplification of the final design, it follows the longitudinal material budget. The protons' position and deposited energy when they traverse each epitaxial layer are stored in ROOT files \cite{brun-root-1997}, the framework in which this analysis is performed.

\subsection{The Proton Beam Setup}\label{sec:protonbeam}
A mono-energetic pencil beam with energy 250 MeV is degraded by 16 cm water, resulting in a spread-out energy spectrum with a mean energy of \SI{180.7}{\MeV}, and an energy spread of \SI{1.4}{\MeV}. The spatial distributions of the pencil beam are described initially by a single Gaussian, with different values of $\sigma_{xy}$ ranging from 2 mm to 5 mm, including a setup with an elliptic beam. The beam divergence is \SIrange{3}{4}{\mrad}, with an emittance of \SIrange{15}{20}{\mm \mrad}, both depending on the direction in $\theta\varphi$-space. See Fig.\ \ref{fig:protonbeam} for an illustration of the asymmetric beam profile. The spatial and angular distributions are expected to gain some non-Gaussian tails due to MCS and elastic nuclear interactions \cite{gottschalk-techniques-2012}.

In order to study the effects of different beam intensities on the track reconstruction properties, the number of protons per pencil beam, $n_p$, is adjusted during the analysis: from $n_p=3$ to $n_p=1000$.



\subsection{The Tracking Algorithm}\label{sec:tracking-intro}
\begin{figure}
\includegraphics[width=\columnwidth]{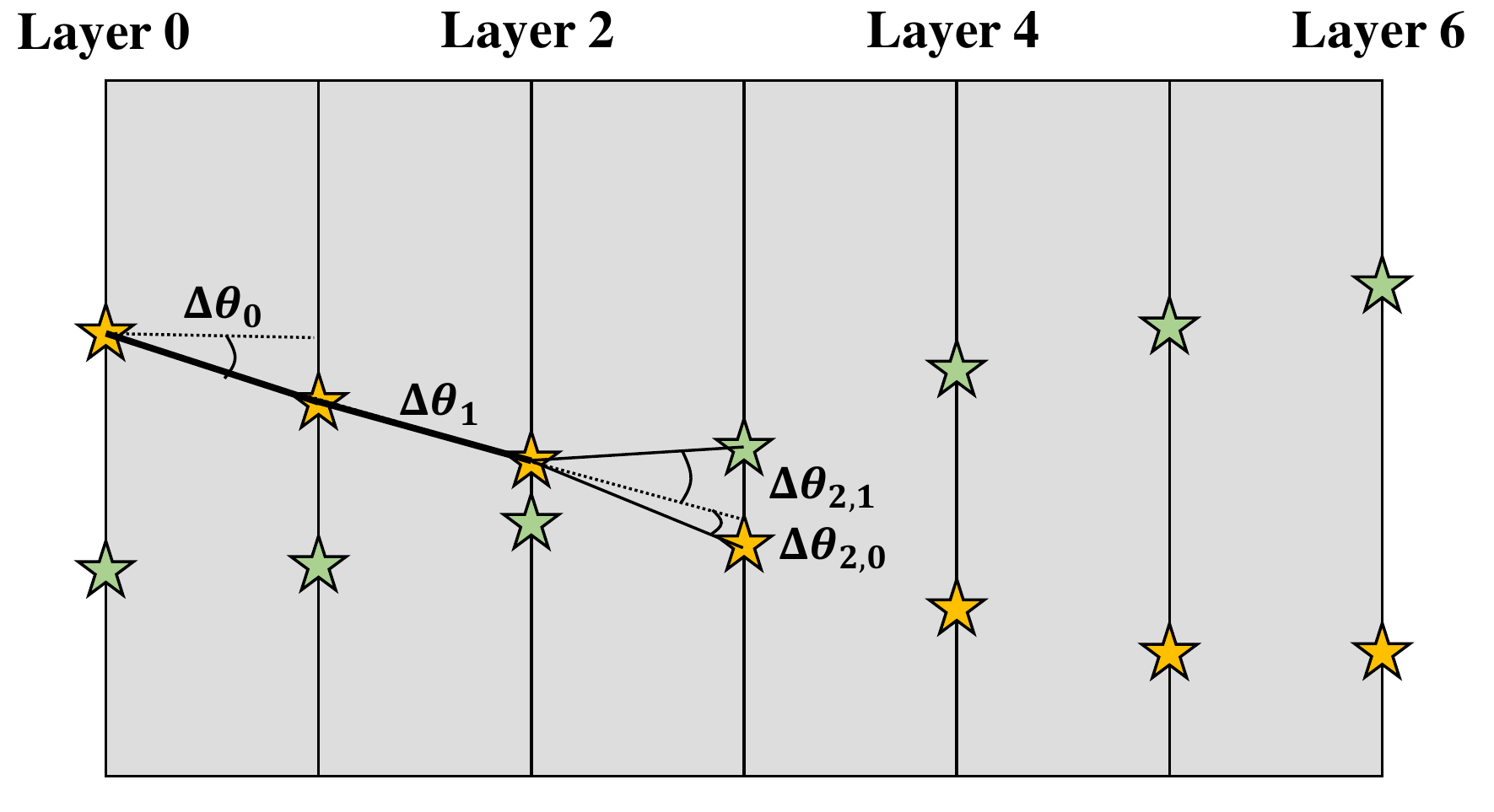}
\caption{Example of the track reconstruction: In this case $\Delta\theta_{2,1}\gg \Delta\theta_{2,0}$ and the latter is chosen as the single next track segment.}\label{fig:tracking}
\end{figure}
An in-detector track reconstruction algorithm needs to be able to handle the proton intensities and the high levels of MCS between subsequent layers that are usually associated with the protons of therapeutic energies (increasing MCS with depth). Elastic nuclear interactions, manifested through infrequent high angle deviations, are also to be expected. The algorithm should be able to discriminate between the different possible endpoints of the protons, such as between Bragg stopping and inelastic nuclear interactions:\footnote{This discrimination is based on the energy deposited in the epitaxial layer, estimated from the size of the charge diffused area of activated pixels around the track. For more details see \cite{pettersen-thesis-2018}.} each incident proton is expected to come to a complete stop inside the detector.

Several strategies for the tracking algorithm have been explored, with the main idea being a "track-following" scheme \cite{strandlie-track-2010}. Some of the ideas behind the algorithm, such as calculating a weight for each hit, based on the angular change of the track, $S$, and comparing the accumulated weight to against a global maximum, \smax, are described in \cite{amrouche-track-2017} as the \textsc{hyperbelle\_tree\_6} solution to a pre-defined tracking challenge. In \cite{pettersen-proton-2017,pettersen-thesis-2018} an earlier version of this algorithm was briefly described, however there it was only applied on a broad and uniform irradiation field, and with inferior performance. The procedure here adapted is described below, with reference to Fig.\ \ref{fig:tracking}:
\begin{enumerate}[i)]
\item Identify all seed pairs in the first two layers, accounting for large incoming angles.

\item Find the angular change $\Delta \theta$ for each seed pair, defined as the change in the 3D vector going in to and out from a layer, and calculate
\begin{math}
S_n = \sqrt{\sum_{\mathrm{layer}}^n(\Delta\theta_{\mathrm{layer}})^2}
\end{math}. In the first layer, a parallel incident beam is assumed.

\item For each seed pair, identify hits in the next layer where $S_{n+1} < S_{\mathrm{max}}$. If several such hits are identified, the best is chosen as the next track segment. Both are chosen if two hits yield sufficiently similar $S_{n+1}$ values.\footnote{Too many forks in the track reconstruction lead to a very slow procedure, due to the many layers and high proton densities involved (there are $\sim 2^{40}$ possibilities if the track is split at each layer). To counteract this, an additional candidate in a layer is included only if both candidates have scores within 15\% of each other.}

In order to avoid excluding straight segments of tracks with $S_n$ values close to \smax{}, track candidates where $\Delta \theta_{n+1} < 50\ \si{\mrad}$ are always allowed even if $S_{n+1} > S_{\mathrm{max}}$. In Section \ref{sec:optimalsmax} this limit will be found.

\item Repeat the above step and and follow all track candidates in the "tree" recursively. The final track with the lowest $S_n$ score from a single seed is kept, and its hits are removed from the search pool.
\end{enumerate}
\noindent
Since the angular distribution of the incoming beam is more parallel at the detector entrance, compared to at the stopping position, and since the protons usually do not stop at the same depth, the track-following is initiated at the front face. However, some improvements may be made here---see the discussion.

The value of \smax{} is adjusted as to yield the highest fraction of correctly reconstructed tracks.

The implementation of this tracking algorithm, as well as the GATE macro files used for the creation of the MC simulations, are available at GitHub \cite{pettersen-github-dtctoolkit}.

\begin{figure}
\includegraphics[width=\columnwidth]{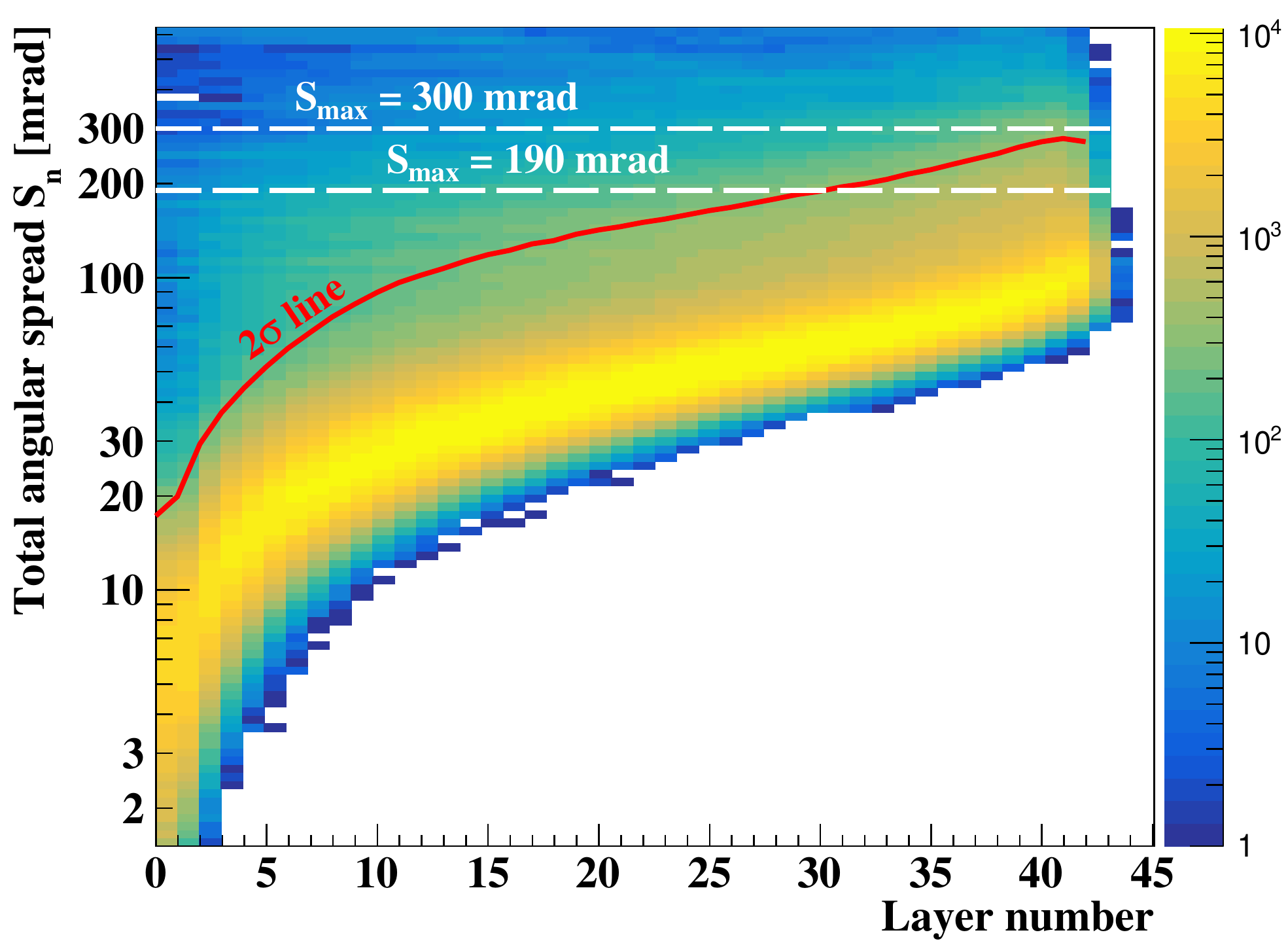}
\caption{Distribution of actual $S_n = \sqrt{\sum_{\mathrm{layer}}^n(\Delta\theta_{\mathrm{layer}})^2}$ values from tracks reconstructed using MC truth. The empirical $2\sigma$ value of the distribution as well as two different \smax{} thresholds are shown.} \label{fig:s-values}
\end{figure}

\subsection{Evaluation of the Tracking Algorithm}\label{sec:evaluationoftracking}
The quantitative evaluation of the tracking algorithm is based on a comparison between the reconstructed tracks, using the current algorithm, and the true tracks from MC simulations. The tracking efficiency is calculated at various pencil beam densities, by finding the fraction of correctly reconstructed tracks.

Prior to the evaluation, a filtering scheme is applied: tracks with higher incoming angles than $3\sigma$ of the distribution of all angles, and less than $3\sigma$ of the range are cut away. In addition, in order to further clean incorrectly reconstructed tracks, a cut on the deposited energy (in terms of the number of pixels activated in the cluster) for protons not stopping in a Bragg peak has been applied---see Fig.\ 4.16 in \cite{pettersen-thesis-2018} for more details. In all, roughly 30--40\% of the protons are removed from the analysis, most of these due to unavoidable physical interactions such as (in)elastic collisions \cite{johnson-review-2018}.

A correctly reconstructed track is defined as following the same primary proton from GATE (\texttt{eventID}) at its start- and endpoints, and it has to be fully tracked. On the other hand, if a secondary particle is tracked and identified as a primary, if two tracks are confused or if a simulated proton continues beyond the reconstructed track, it is counted as a \textit{fake track} and is not correctly reconstructed.

\subsection{Finding the Optimal S\textsubscript{max} Value} \label{sec:optimalsmax}
The choice of \smax{} determines the amount of scattering that is allowed for a given track. Too small values lead to prematurely discarded track candidates, and too large values cause confusion by including wrong candidates where there should be none, \textit{e.g.}\ after inelastic nuclear interactions at high particle densities.

In Fig.\ \ref{fig:s-values} a two-dimensional histogram is shown of the distribution of $S_n$ values in each layer, found using true tracks from MC simulations. The summed angular spread $S_n$ is distributed with long tails, resembling long-tailed Landau distributions in the deeper layers. The $2\sigma$ value in the layer where most of the protons stop is \SI{270}{\mrad} (determined by vertically summing the bin areas up to 97.7\%).

Analytically, the Highland equation (in three dimensions) as given in \cite{gottschalk-scattering-2010} gives the $2\sigma$ value of the $\Delta \theta$ scattering angle:
\begin{equation}
2\sigma_{\Delta \theta} = 2\sqrt{2} \left( \int_0^x \left( \frac{14.1\ \mathrm{MeV}}{pv(x')}\right)^2 \frac{1}{X_0}\mathrm{d}x' \right)^{1/2} \left(1 + \frac{1}{9} \log_{10} \frac{x}{X_0}\right), \label{eq:highland}
\end{equation}
In this case, the radiation length $X_0 = \SI{77.9}{mm}$ is found by adding the various materials in Fig.\ \ref{fig:layer-setup}, and $pv$ are the momentum and velocity factors. By performing a numerical integration up to $x=0.95R$ (for consistency with the ranges in Fig.\ \ref{fig:s-values}), we find that $2\sigma_{\Delta \theta} = \SI{278}{\mrad}$.

However, while this value can be applied as the \smax{} in the track reconstruction algorithm on a track-by-track basis, it is not readily apparent how to account for high particle densities. Large values of \smax{} might allow for the incorrect continuation of tracks ending due to inelastic collisions, by following nearby tracks.

\begin{figure}
\includegraphics[width=\columnwidth]{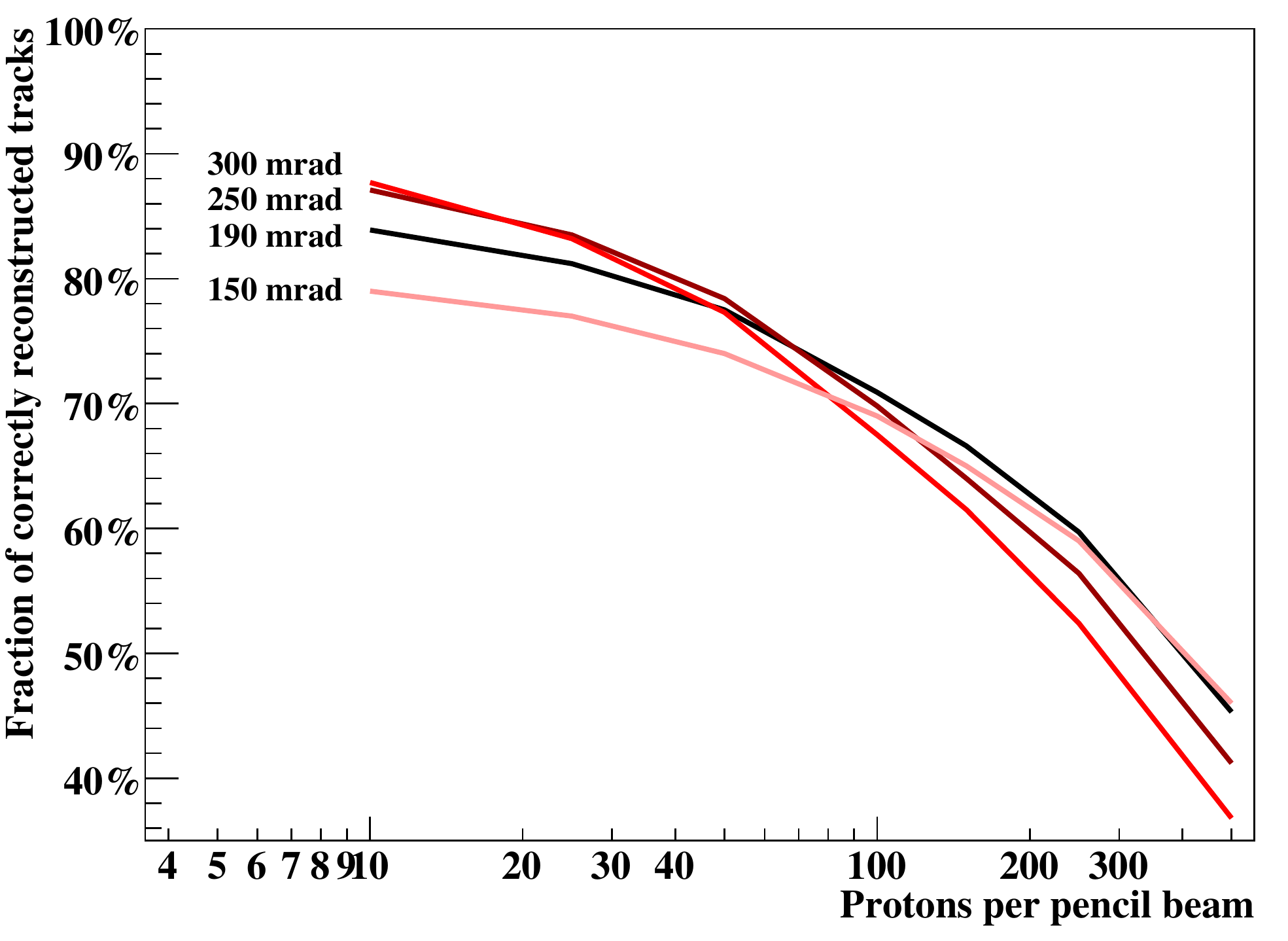}
\caption{The reconstruction efficiency of a pencil beam of increasing density, evaluated using different \smax{} thresholds (labeled in figure). Notice that each of the \smax{} curves yields the maximum efficiency at a certain density range.}\label{fig:smax-vs-efficiency}
\end{figure}

In the following, in order to account for this density effect, the optimal \smax{} for this algorithm is found by identifying the \smax{} values that yield the lowest fake rate at increasing densities.\footnote{Note that in order to find the best \smax{} parameter, no $3\sigma$ filtering on the range is applied here---a filter that reduces the fake rate but also hides many of the tracks that fail to reconstruct properly.} This function is found from a parameter scan (see Fig.\ \ref{fig:smax-vs-efficiency}), and can be parametrized as
\begin{align}
S_{\mathrm{max}}(n_p) &= 470\ n_p^{\ \ -0.176}\ \si{\mrad}.
\end{align}
Here the beam spot size is \SI{3}{\mm}, similar expressions are found for different spot sizes.

\section{Results}\label{sec:results}
\subsection{Examples of the Track Reconstruction}
\begin{figure}
\includegraphics[clip, trim=0.2cm 0.2cm 2.5cm 1.9cm, width=\columnwidth]{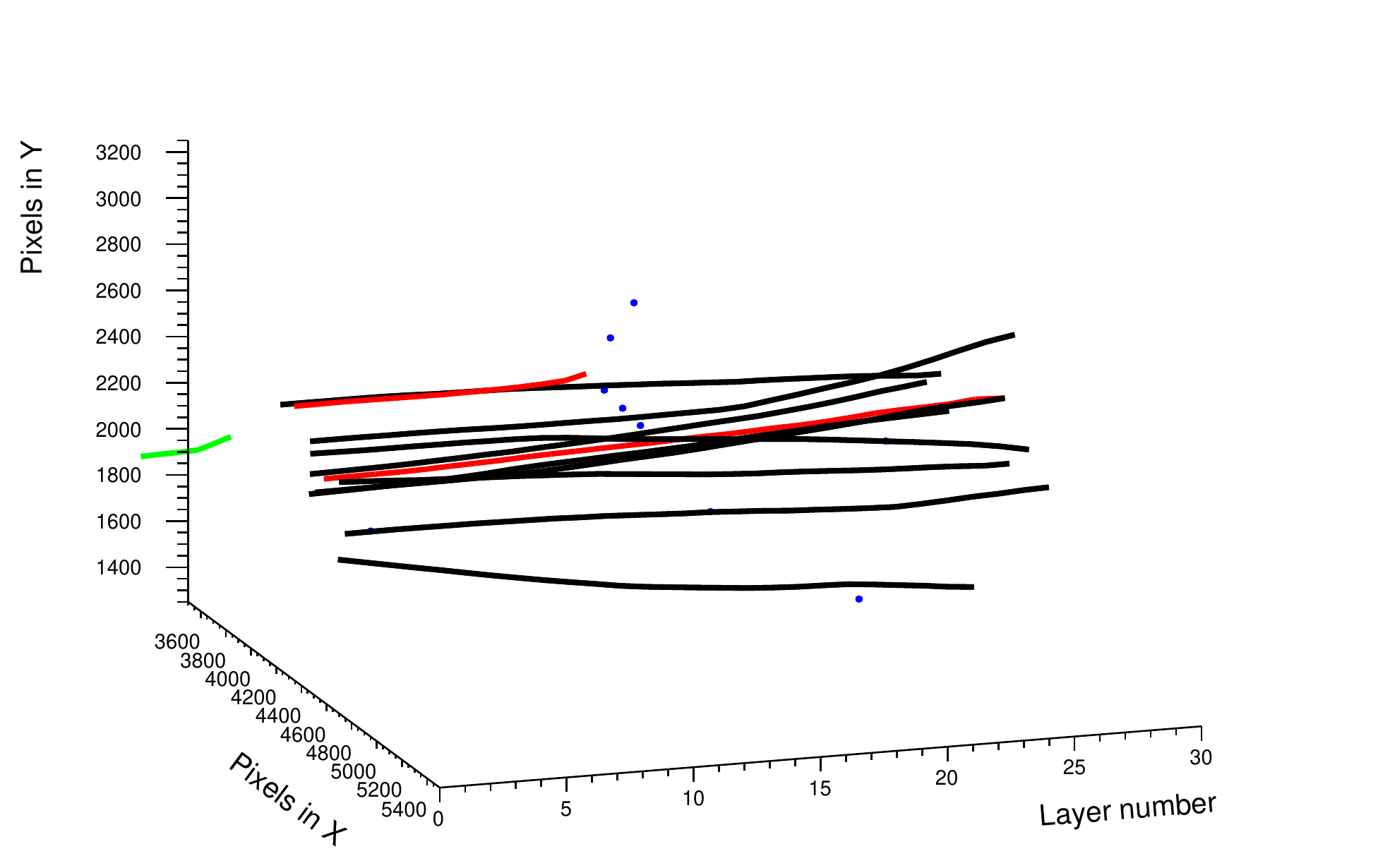}
\caption{Examples of several reconstructed tracks. Correctly reconstructed tracks are visualized as black lines, incorrectly reconstructed fake tracks as red lines and unused pixel hits as blue dots. Green lines are reconstructed secondary particles (here the high angle of the visible secondary particle pushes it out of the view frame). }\label{fig:reconstructedtracks}
\end{figure}

If the reconstruction is configured with the \smax{} values as suggested in the last section, the efficiency should be as high as attainable from the algorithm described in Section \ref{sec:tracking-intro}. An example of the track reconstruction applied on the simulated output from a pencil beam with $\sigma_{xy} = \SI{3}{\mm}$ is shown in Fig.\ \ref{fig:reconstructedtracks}. In the figure, 13 of the 16 tracks have been reconstructed correctly, the remainder involving inelastic nuclear interactions or confusion due to MCS.

\subsection{Tracking Efficiency with Increasing Particle Density}
To quantify the effect of the particle density on reconstruction efficiency, a set of reconstructions with increasing $n_p$ was performed. Several reconstructions are averaged to reach the same number of particle tracks (at least \num{5000}).

By repeated reconstruction of $n_p = 100$ initial protons\footnote{Note that \SI{\sim 8}{\percent} are lost inside the water phantom.}, 77\% of them are found to be reconstructed correctly, according to the criteria outlaid in Section \ref{sec:evaluationoftracking}. From the filtering, 12\% are removed due to $>3\sigma$ incoming angles, 12\% again from the cut on the deposited energy at the track's end and 2\%--3\% due to $<3\sigma$ range. Visually, it can be deduced that the remaining fake tracks exhibit large angle scattering (due to (in)elastic nuclear collisions) inside the tracking detector, or are located in the high-density Gaussian core of the pencil beam where it is easy to confuse tracks.

\begin{figure}
\includegraphics[width=\columnwidth]{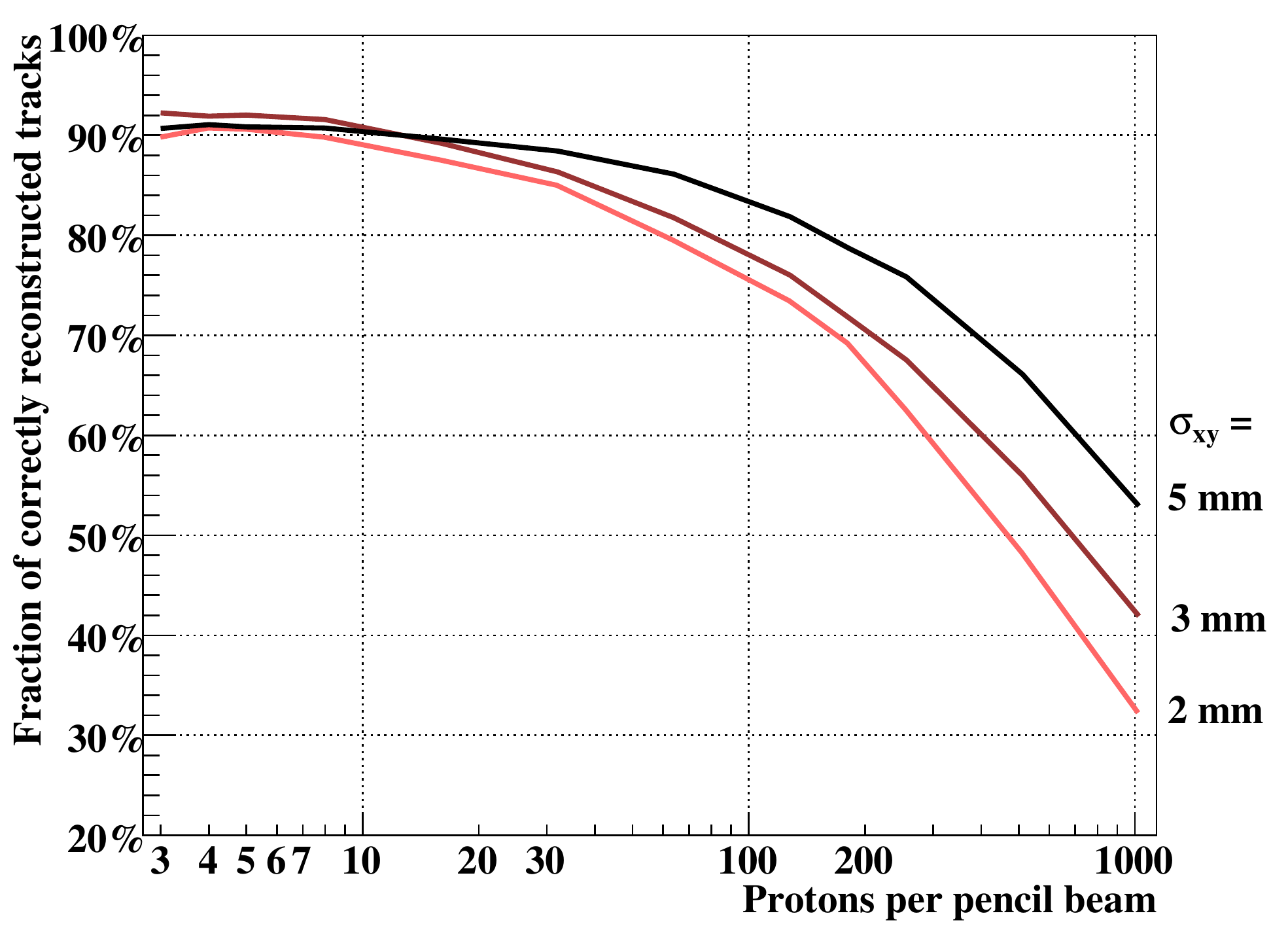}
\caption{Fraction of correctly reconstructed tracks in three pencil beams with different circular spot sizes of $\sigma_{xy} =$ 2--5 mm.} \label{fig:results_pencilbeam}
\end{figure}

\begin{figure}
\includegraphics[width=\columnwidth]{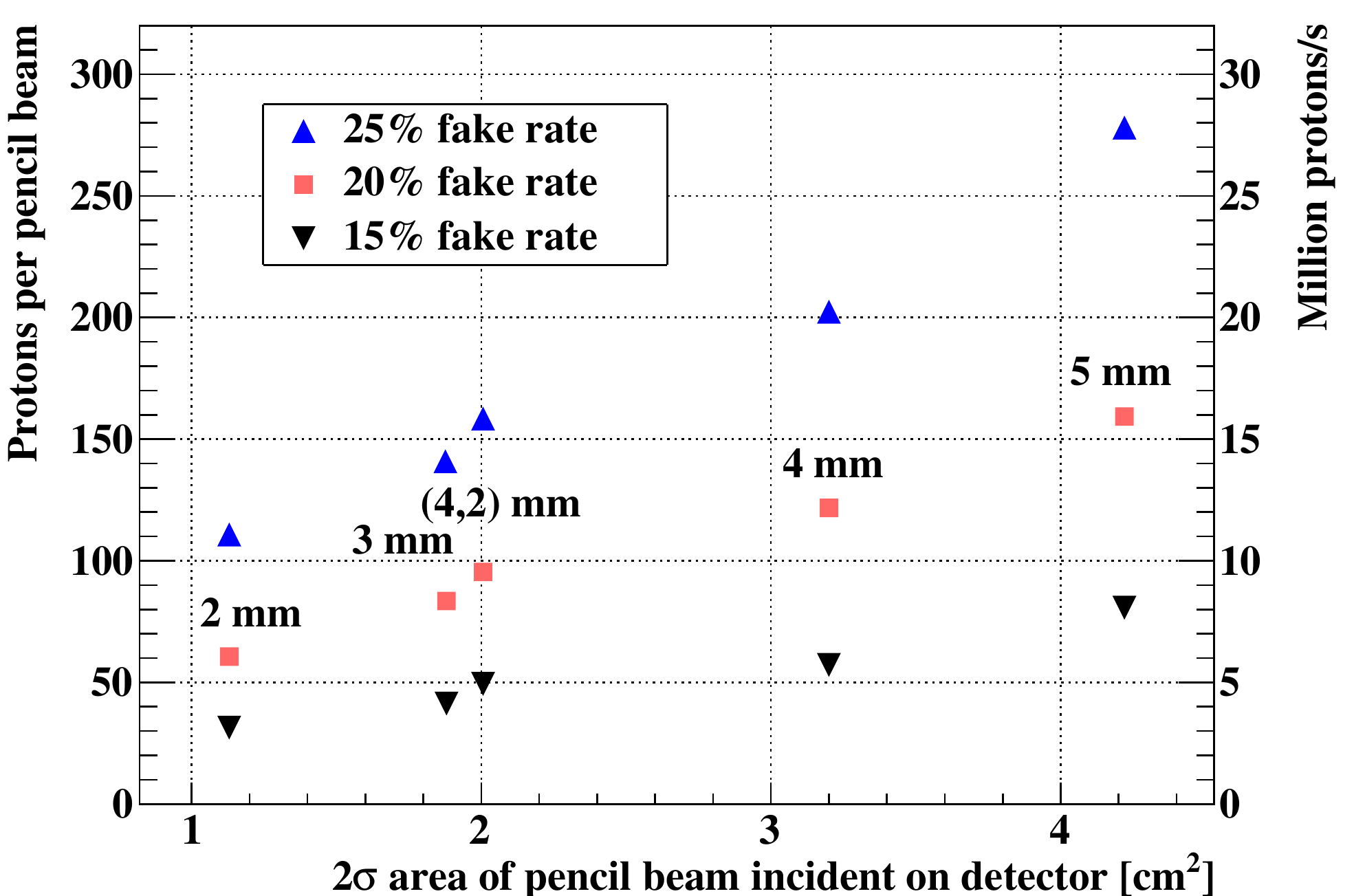}
\caption{Proton Intensity Capacity at 15\%--25\% fake rates, for different beam spot sizes in the range \SIrange{2}{5}{\mm} (including the asymmetric beam). The number $n_p$ per reconstruction frame has also been converted to $n_p$ per second using the expected pixel telescope readout frequency of \SI{100}{\kHz} \cite{pettersen-thesis-2018,mager-alpide-2016}.}\label{fig:capacity-vs-2sigma}
\end{figure}

In Fig.\ \ref{fig:results_pencilbeam} the results from this procedure are presented, for several pencil beams of different circular spot sizes. It can be seen in Fig.\ \ref{fig:capacity-vs-2sigma} that the $n_p$ yielding, respectively, 80\% or 75\%  efficiency increases linearly with the beam area for different beam spot sizes.

\subsection{Effects of Scattering on Tracking Efficiency} \label{sec:differentphysics}
The efficiency of the tracking algorithm depends upon the different interactions a particle may undergo. The amount of MCS depends on the traversed material, and there is a certain probability that the particle undergoes elastic or inelastic nuclear interactions.

\begin{figure}
\includegraphics[width=\columnwidth]{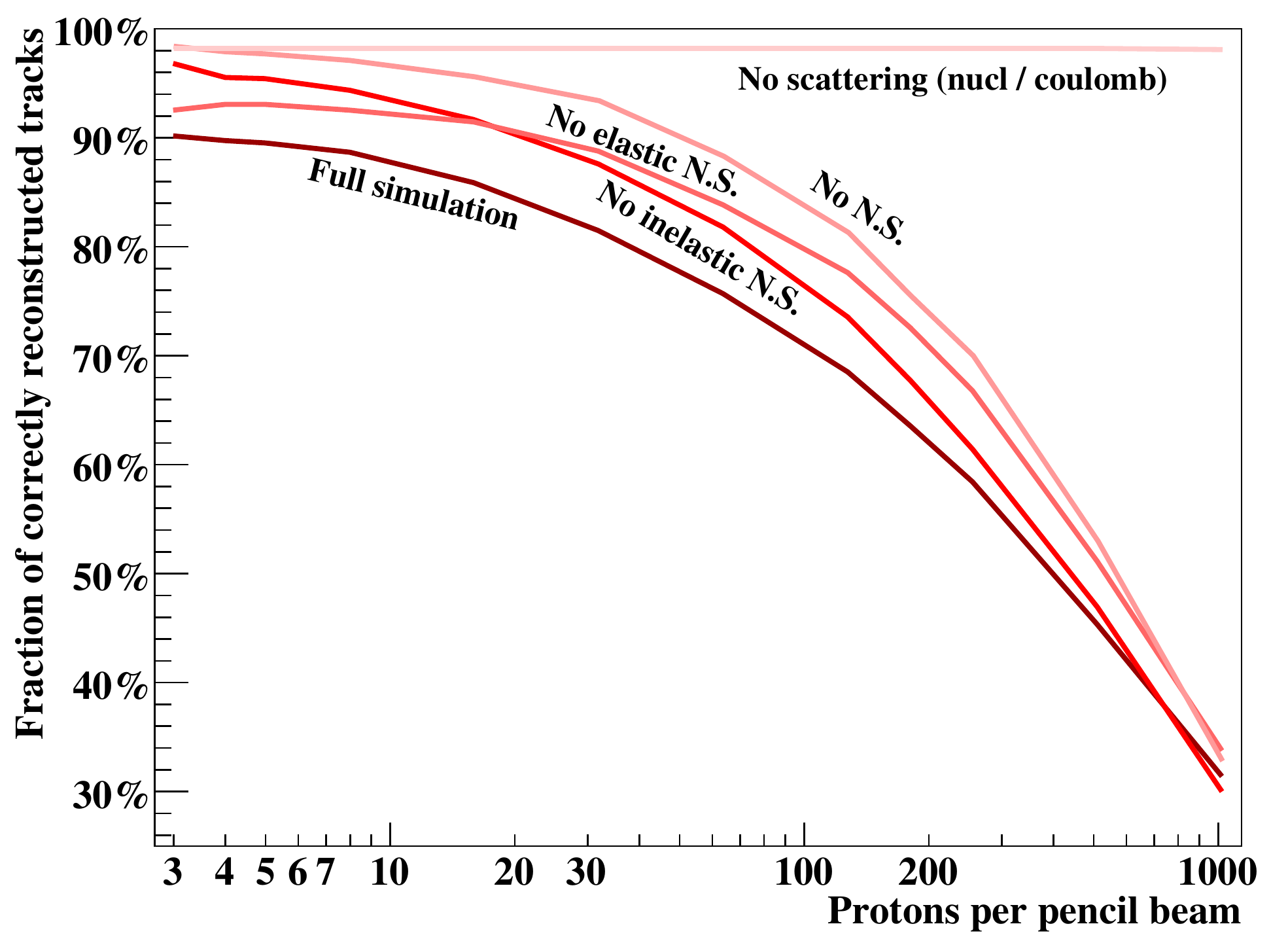}
\caption{The fraction of correctly reconstructed tracks when different interactions between the target protons and the detector materials are turned on or off in the MC simulations. The interactions are inelastic / elastic nuclear scattering (N.\ S.) and multiple Coulomb scattering.} \label{fig:efficiency-different-physics-mcs}
\end{figure}

To quantify the effects of these different interactions on the quality of the track reconstruction, a set of simulations were performed, including or excluding these effects. The simulations were performed with the $\sigma_{xy} = \SI{3}{\mm}$ beam.

In Fig.\ \ref{fig:efficiency-different-physics-mcs} five scenarios are outlaid: Using the standard settings of \texttt{QGSP\_BIC\_EMZ} in a "full simulation"; elastic nuclear scattering switched off; inelastic nuclear scattering switched off; all nuclear scattering switched off; and all scattering switched off (including MCS). The following features are seen:
\begin{itemize}
\item At low densities particle densities, the most degrading interaction type is the inelastic nuclear scattering. A constant fraction (approx.\ 1\% per water equivalent cm \cite{durante-nuclear-2016}) of the protons undergo inelastic scattering. The primary particle is lost and a subsequent search for its continuation may lead to false candidates---especially is \smax{} is high. Approximately 70\%  of the fake tracks are due to secondary particles at $n_p < 5$.

\item At high particle densities, MCS causes increasing amounts of confusion in the track reconstruction process, and as a result this effect is the most degrading at densities where $n_p > 50$. When MCS is turned off, the track reconstruction is trivial and the remaining constant fake rate of 1\%--2\% is due to confusion from delta rays.
\end{itemize}

\subsection{Computational Demands}
The full track reconstruction as described here has been implemented in C++ / ROOT6 \cite{brun-root-1997} for single CPU usage. Several reconstruction batches can be run in parallel with small memory footprint. The reconstruction time per incoming proton is \SIrange{2}{6}{\ms}, depending on the overhead (an important contribution to the total time for $n_p < 100$) and proton density.

While the code has been optimized, further optimization and parallelization is needed if the required $\sim 10^{8}$ protons are to be tracked within 5--10 minutes.

\section{Discussion}
The tracking algorithm as presented in this study shows promise for use in a pixel-based range telescope for proton imaging. Further improvements of the tracking algorithm, apart from reducing computational requirements, might include the optimization of the order in which the tracks are reconstructed (starting with low-density areas of the pencil beam), bidirectional reconstruction (based on cellular automaton) to better classify large angle scattering and to reduce confusion arising from protons stopping in different layers \cite{strandlie-track-2010}, identifying and connecting track segments using concepts from graph theory \cite{sikler-combination-2017} and even novel methods based on deep neural networks \cite{farrell-hep.trkx-2017}. While reasons for performing the reconstruction starting from the front face of the detector have been outlaid, it would be interesting to try selecting seeds from the distal layers of the detector, in which case both the stopping- and starting positions of each track would be known.

The obtained results, representing the theoretical limits of the current algorithm, reflect the high amounts of scattering inherent in a proton beam at therapeutic energies and spatial distributions. This limitation can be reduced if heavier ions are used for the imaging process, as they exhibit less scattering \cite{durante-nuclear-2016}. On the other hand, if heavier ions are to be applied, an improved identification of particle species of the mixed beam is necessary due to the increased projectile fragmentation. The pixel matrix, together with the possibility of calculating the deposited energy in each layer, enables such identification of different physics processes in the detector.

The beam intensity capabilities have been calculated using fake rates of the reconstruction of 15\%--25\%. No comprehensive study of the optimal target fake rate has been performed, and it is expected that fake tracks can be further filtered out by applying $3\sigma$ filter on the WEPL values in the angle / position bin during the reconstruction process \cite{schulte-density-2005}. The determination of the maximum allowed fake rate can thus be made from constraints such as minimizing the radiation dose to patient.

\section{Conclusions}\label{conclusions}
In this work a tracking algorithm for protons traversing a pixel-based range telescope has been presented. The performance of the algorithm on a simulated setup has been shown to be of sufficient quality, in terms of the maximum proton beam intensity that can be reconstructed simultaneously --- in the order of 5--25 million protons/s (depending on spot size and fake rate). Some proton CT requirements, such as spatial resolution in the reconstructed volumes, are best fulfilled using the smaller beam spot sizes \cite{krah-2018-comprehensive}.

The tracking algorithm presented here enables a pixel-based range telescope setup to reconstruct the incoming angles and final ranges of a large number of concurrent proton tracks. For a proton CT system based upon this to enter into the clinic, a better performing reconstruction algorithm is required, particularly in terms of reconstruction time. The track reconstruction framework has been developed and implemented through a productive knowledge transfer from High Energy Physics efforts.

\section{Acknowledgements}\label{sec:acknowledgements}
This project has been supported by Helse Vest RHF grant [911933] and Bergen Research Foundation grant [BFS2015PAR03].

\bibliography{ptracking}

\begin{thebibliography}{27}

\bibitem{jermann-particle-2017}
M.~Jermann, PTCOG  (2017)

\bibitem{dionisi-use-2014}
F.~Dionisi, E.~Ben-Josef, The Cancer Journal \textbf{20} (2014)

\bibitem{leeman-proton-2017}
J.E. Leeman, P.B. Romesser, Y.~Zhou, S.~McBride, N.~Riaz, E.~Sherman, M.A.
  Cohen, O.~Cahlon, N.~Lee, The Lancet Oncology \textbf{18}, e254 (2017)

\bibitem{schneider-calibration-1996}
U.~Schneider, E.~Pedroni, A.~Lomax, Physics in medicine and biology
  \textbf{41}, 111 (1996)

\bibitem{paganetti-range-2012}
H.~Paganetti, Physics in Medicine and Biology \textbf{57}, R99 (2012)

\bibitem{baer-potential-2017}
E.~Bär, A.~Lalonde, G.~Royle, H.M. Lu, H.~Bouchard, Medical Physics
  \textbf{44}, 2332 (2017)

\bibitem{johnson-review-2018}
R.P. Johnson, Reports on Progress in Physics \textbf{81}, 016701 (2018)

\bibitem{krah-2018-comprehensive}
N.~Krah, F.~Khellaf, J.M. Létang, S.~Rit, I.~Rinaldi, Physics in Medicine and
  Biology \textbf{63}, 135013 (2018)

\bibitem{pettersen-thesis-2018}
H.E.S. Pettersen, {PhD thesis}, University of Bergen, Norway (2018)

\bibitem{karbasi-incorporating-2015}
P.~Karbasi, B.~Schultze, V.~Giacometti, T.~Plautz, K.~Schubert, R.~Schulte,
  V.~Bashkirov, 2015 IEEE Nuclear Science Symposium and Medical Imaging
  Conference  (2015)

\bibitem{rit-filtered-2013}
S.~Rit, G.~Dedes, N.~Freud, D.~Sarrut, J.M. Létang, Medical Physics
  \textbf{40}, 031103 (2013)

\bibitem{r.-p.-johnson-fast-2016}
R.P. Johnson, V.~Bashkirov, L.~DeWitt, V.~Giacometti, R.F. Hurley, P.~P., T.E.
  Plautz, H.F.W. Sadrozinski, K.~Schubert, R.~Schulte et~al., IEEE Transactions
  on Nuclear Science \textbf{63}, 52 (2016)

\bibitem{pettersen-proton-2017}
H.~Pettersen, J.~Alme, A.~Biegun, A.~van~den Brink, M.~Chaar, D.~Fehlker,
  I.~Meric, O.~Odland, T.~Peitzmann, E.~Rocco et~al., Nuclear Instruments and
  Methods in Physics Research Section A: Accelerators, Spectrometers, Detectors
  and Associated Equipment \textbf{860}, 51 (2017)

\bibitem{mager-alpide-2016}
M.~Mager, Nuclear Instruments and Methods in Physics Research Section A:
  Accelerators, Spectrometers, Detectors and Associated Equipment \textbf{824},
  434 (2016)

\bibitem{esposito-cmos-2015}
M.~Esposito, T.~Anaxagoras, P.~Evans, S.~Green, S.~Manolopoulos,
  J.~Nieto-Camero, D.~Parker, G.~Poludniowski, T.~Price, C.~Waltham et~al.,
  Journal of Instrumentation \textbf{10}, C06001 (2015)

\bibitem{amrouche-track-2017}
S.~Amrouche, N.~Braun, P.~Calafiura, S.~Farrell, J.~Gemmler, C.~Germain, V.V.
  Gligorov, T.~Golling, H.~Gray, I.~Guyon et~al., \emph{Track reconstruction at
  {LHC} as a collaborative data challenge use case with {RAMP}} (EPJ Web of
  Conferences, Paris, France, 2017), Vol. 150

\bibitem{jan-gate:-2004}
S.~Jan, G.~Santin, D.~Strul, S.~Staelens, K.~Assie, D.~Autret, S.~Avner,
  R.~Barbier, M.~Bardies, P.M. Bloomfield et~al., Physics in medicine and
  biology \textbf{49}, 4543 (2004)

\bibitem{grevillot-optimization-2010}
L.~Grevillot, T.~Frisson, N.~Zahra, D.~Bertrand, F.~Stichelbaut, N.~Freud,
  D.~Sarrut, Nuclear Instruments and Methods in Physics Research Section B:
  Beam Interactions with Materials and Atoms \textbf{268}, 3295 (2010)

\bibitem{brun-root-1997}
R.~Brun, F.~Rademakers, Nuclear Instruments and Methods in Physics Research
  Section A: Accelerators, Spectrometers, Detectors and Associated Equipment
  \textbf{389}, 81 (1997)

\bibitem{gottschalk-techniques-2012}
B.~Gottschalk, arXiv preprint arXiv:1204.4470  (2012)

\bibitem{strandlie-track-2010}
A.~Strandlie, R.~Frühwirth, Reviews of Modern Physics \textbf{82}, 1419 (2010)

\bibitem{pettersen-github-dtctoolkit}
H.E.S. Pettersen, \emph{{GitHub} - {DTC} toolkit},
  \url{https://github.com/HelgeEgil/focal} (2015)

\bibitem{gottschalk-scattering-2010}
B.~Gottschalk, Medical Physics \textbf{37}, 352 (2010)

\bibitem{durante-nuclear-2016}
M.~Durante, H.~Paganetti, Reports on Progress in Physics \textbf{79}, 096702
  (2016)

\bibitem{sikler-combination-2017}
F.~Siklér, \emph{Combination of various data analysis techniques for efficient
  track reconstruction in very high multiplicity events} (EPJ Web of
  Conferences, Paris, France, 2017), Vol. 150

\bibitem{farrell-hep.trkx-2017}
S.~Farrell, D.~Andersen, P.~Calafiura, G.~Cerati, L.~Gray, J.~Kowalkowski,
  M.~Mudigonda, Prabhat, P.~Spentzouris, M.~Spiropoulou et~al., \emph{The
  {HEP}.{TrkX} {Project}: deep neural networks for {HL}-{LHC} online and
  offline tracking} (EPJ Web of Conferences, Paris, France, 2017), Vol. 150

\bibitem{schulte-density-2005}
R.W. Schulte, V.~Bashkirov, M.C. Loss~Klock, T.~Li, A.J. Wroe, I.~Evseev, D.C.
  Williams, T.~Satogata, Medical Physics \textbf{32}, 1035 (2005)

\end{thebibliography}

\end{document}